\documentclass[%
reprint,
amsmath,amssymb, 
aps,
prb
]{revtex4-2}
\usepackage{graphicx, color}
\usepackage{dcolumn}
\usepackage{bm}
\usepackage[colorlinks,urlcolor=blue,linkcolor=blue,anchorcolor=blue,citecolor=blue]{hyperref} 

\begin{document}
\preprint{APS/123-QED}
\title{Type-II  Dirac nodal chain semimetal CrB$_4$}
\author{Xiao-Yao Hou$^{1,2}$}
\author{Ze-Feng Gao$^{1,2}$}
\author{Peng-Jie Guo$^{1,2}$}
\email{guopengjie@ruc.edu.cn}
\author{Jian-Feng Zhang$^{3}$}
\email{zjf@iphy.ac.cn}
\author{Zhong-Yi Lu$^{1, 2, 4}$}
\email{zlu@ruc.edu.cn}
\affiliation{1. School of Physics and Beijing Key Laboratory of Opto-electronic Functional Materials $\&$ Micro-nano Devices. Renmin University of China, Beijing 100872, China}
\affiliation{2. Key Laboratory of Quantum State Construction and Manipulation (Ministry of Education), Renmin University of China, Beijing 100872, China}
\affiliation{3. Institute of Physics, Chinese Academy of Sciences, Beijing 100190, China}
\affiliation{4. Hefei National Laboratory, Hefei 230088, China}


\date{\today}
\begin{abstract}
 Dirac nodal line semimetals with topologically protected drumhead surface states have attracted intense theoretical and experimental attention over a decade. However, the study of type-II Dirac nodal line semimetals is rare, especially the type-II nodal chain semimetals have not been confirmed by experiment due to the lack of ideal material platform. In this study, based on symmetry analysis and the first-principles electronic structure calculations, we predict that CrB$_4$ is an ideal type-II Dirac nodal chain semimetal protected by the mirror symmetry. Moreover, there are two nodal rings protected by both space-inversion and time-reversal symmetries in CrB$_4$. More importantly, in CrB$_4$ the topologically protected drumhead surface states span the entire Brillouin zone at the Fermi level. On the other hand, the electron-hole compensation and the high mobility of Dirac fermions can result in large magnetoresistance effects in CrB$_4$ according to the two-band model. Considering the fact that CrB$_4$ has been synthesized experimentally and the spin-orbit coupling is very weak, CrB$_4$ provides an ideal material platform for studying the exotic properties of type-II Dirac nodal chain semimetals in experiment.

\end{abstract}

\maketitle

\section{Introduction}

Topological semimetals with symmetry-protected band crossing  have attracted intense theoretical and experimental interest due to numerous novel physical properties, such as topologically protected boundary state, chiral anomaly, chiral zero sound, large magnetoresistance effect, topological catalysis\cite{Weng_2016, novel-1, novel-2, novel-3, Li-apl, xie-sm}. According to both the degeneracy and dimension of band crossing, topological semimetals are divided into zero-dimensional Weyl semimetals\cite{wan-2011,Weng-prx}, Dirac semimetals\cite{2-Dirac1,2-Dirac2,2-Dirac3, Na3Bi}, triple degenerate semimetals\cite{science-fermions, triply-1, triply-2, Guo-triple}, sixfold degenerate semimetals\cite{science-fermions,Sixfold-1, Sixfold-2}, eightfold degenerate semimetals\cite{eightfold-1,science-fermions,eightfold-2}, one-dimensional nodal-line semimetals\cite{Weyl-node-line, Dirac-node-line, yu-prl}, and two-dimensional nodal surface semimetals\cite{wu-prb}. On the other hand, based on whether the zero-dimensional band crossing has strong tilt along a certain direction, zero-dimensional topological semimetals can also be divided into type-I and type-II topological semimetals, such as type-II Weyl\cite{Type-II, Duanprb, type2dirac-1}. In comparison, type-II topological semimetals have unique physical properties different from type-I topological semimetals such as angle-dependent chiral anomaly and topological Lifshitz transitions\cite{Type-II, Duanprb, type2dirac-1}.

One-dimensional nodal line semimetals are divided into Dirac and Weyl nodal line semimetals\cite{Weyl-node-line, Dirac-node-line, yu-prl}. Moreover, multiple nodal lines can also form nodal chain semimetals, hopf-links nodal line semimetals and so on\cite{Bzdusek-IrF4,Hopf-1}. Naturally, the band crossing on the nodal lines can contain type-II Dirac and Weyl points, and the corresponding topological semimetals are called type-II Dirac and Weyl nodal line semimetals respectively. Although several type-I nodal line and nodal chain semimetals have been experimentally confirmed\cite{TiB2-1, ZrB2, nodal-chain}, type-II nodal line semimetals have rarely been experimentally confirmed due to the lack of an ideal material platform. Until recently, ZrSiSe was confirmed as a type-II nodal line semimetal by angle-resolved photoemission spectroscopy (ARPES) measurement\cite{Type-II-Nodal-Line}. However, type-II nodal chain semimetals have yet not been experimentally confirmed. Therefore, it is very meaningful to predict type-II nodal chain semimetals with distinct properties for the study of their novel physical properties in experiment. 

In this study, based on symmetry analysis and the first-principles electronic structure calculations, we predict that CrB$_4$ is an ideal type-II Dirac nodal chain topological semimetal protected by the mirrors M$_z$ and M$_y$ (1/2, 1/2, 1/2). In addition, there are also two nodal rings protected by both space-inversion (I) and time-reversal (T) symmetries. More importantly, the topologically protected Fermi arcs of CrB$_4$ at the Fermi level span the entire Brillouin zone. On the other hand, electron-hole compensation and their high mobility can cause CrB$_4$ to have large magnetoresistance effect. When spin-orbit coupling (SOC) is considered, CrB$_4$ transforms from type-II nodal chain semimetal to strong topological insulator with Z$_2$ topological invariant (1,000).

\section{Method}

The structural optimization and electronic structure calculation of CrB$_4$ were studied in the framework of density functional theory (DFT) \cite{DFT-1,DFT-2}in the Vienna ab initio simulation package (VASP) \cite{VASP-1,VASP-2,VASP-3}. The core electrons as well as the interaction between the core and valence electrons were described by using the projector augmented-wave meth \cite{PAW}. The generalized gradient approximation (GGA) of Perdew-Burke-Ernzerhof (PBE)\cite{PBE} type was treated with exchange-correlation functional, using an energy cut-off of 600 eV for the plane waves. A 14 × 16 × 20 Monkhorst-Pack k mesh was used for the Brillouin zone sampling of the unit cell. The internal atomic positions were fully relaxed until the forces on all atoms were smaller than 0.01 eV/Å. The tight-binding Hamiltonian was constructed by the maximally localized Wannier functions \cite{WannierFunction,W90-1,w90-2} and the corresponding topological properties were obtained by using the WannierTools package \cite{WannierTools}. 

\section{Results}

Chromium tetraboride (CrB$_4$), originally suggested to have the \textit{Immm} space group (No. 71) with the\textit{ oI}10 unit cell in Pearson notation, was known for excellent adhesive wear resistance due to its high hardness \cite{CrB4-Immm}. Afterwards, Niu et al. \cite{CrB4-Pnnm} experimentally confirmed the existence of orthorhombic space group \textit{Pnnm} (No. 58) CrB$_4$ with \textit{op}10 unit cell in Pearson notation, which has a lower symmetry than the former. We optimize the experimental structure for CrB$_4$ with the orthorhombic \textit{Pnnm} space group. The optimized lattice constants are \textit{a} = 4.72 Å, \textit{b} = 5.47 Å, and \textit{c }= 2.85 Å. Due to 3d-orbital electrons with strong correlation, CrB$_4$ may have magnetism. We consider different magnetic structures for CrB$_4$, all of which are optimized to nonmagnetic state. CrB$_4$ is thus a paramagnetic material, which is consistent with the experimental results [49]. The crystal structure of CrB$_4$ is shown in Fig. 1(a-c). From the $a$ direction, CrB$_4$ is composed of the Cr atomic layer and the B atomic layer, while the B atomic layer is formed by the distorted hexagon of B atoms (Fig. 1(a) and (c)), which results in the low symmetry of CrB$_4$. The point group symmetry of CrB$_4$ is D$_{2h}$ with generator elements C$_{2z}$, C$_{2x}$(1/2, 1/2, 1/2) and I. The bulk and  surface Brillouin zone (BZ) with the high-symmetry points and lines are shown in Fig. 1(d).

\begin{figure}
\centering
\includegraphics[width=8.5cm]{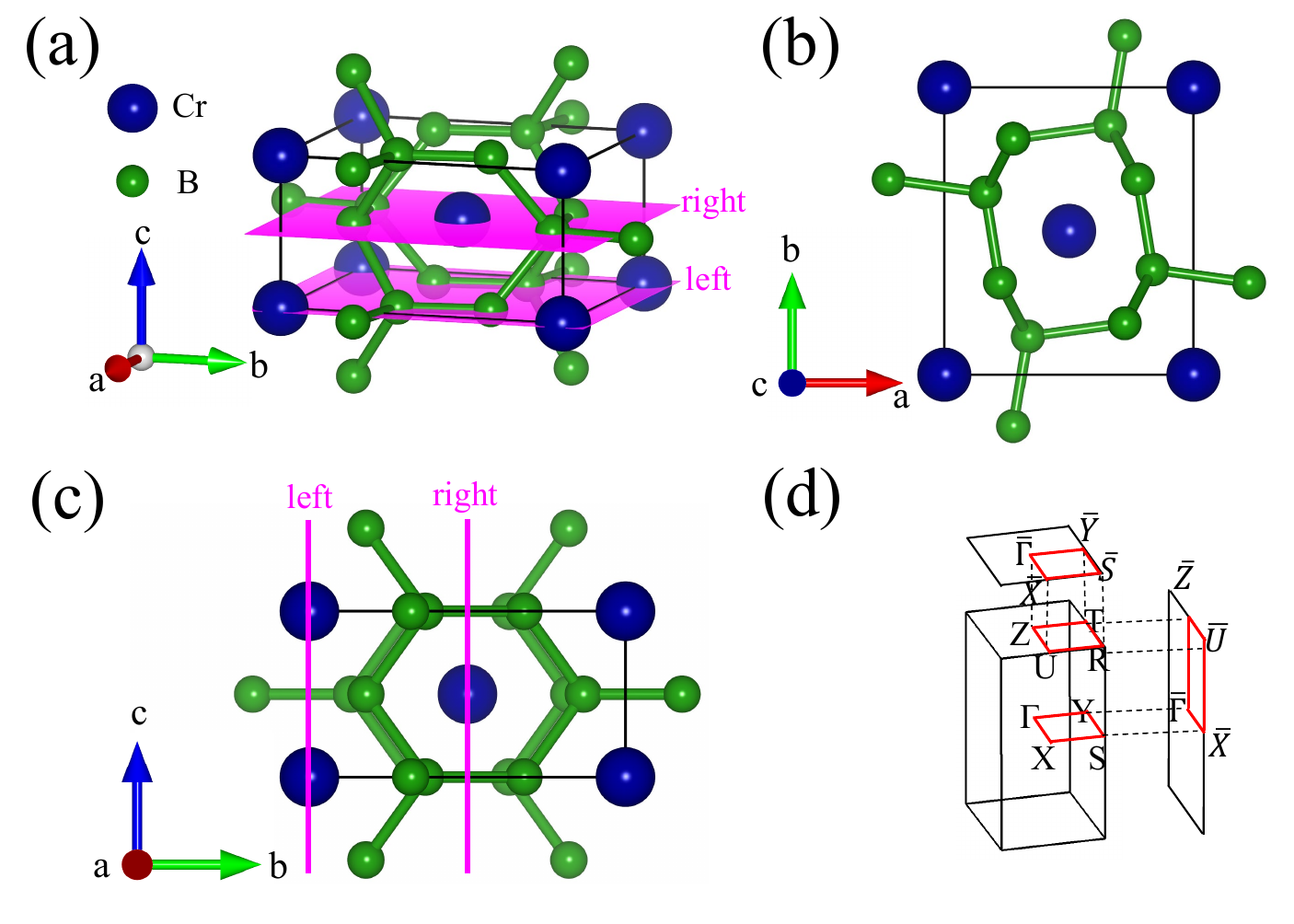}
\caption{\label{fig:1} (a), (b), and (c) are the crystal structures of CrB$_4$ with side view, top view along the c and a directions, respectively. The magenta planes and lines represent the corresponding left and right terminations.(d) Bulk, (001), and (010) projected 2D surface BZs of CrB$_4$, where the high-symmetry lines and the high-symmetry points are marked.}
\end{figure}

To study the topological properties of CrB$_4$, we calculate its electronic band structure along the high-symmetry directions without spin-orbit coupling (SOC), which is shown in Fig. 2(a). From Fig. 2(a), CrB$_4$ is a semimetal and has only blue and yellow bands crossing the Fermi level. Moreover, there are band crossings around the Fermi level, thus CrB$_4$ may be topological semimetal protected by crystal symmetry. Symmetry analysis indicates the band crossing points at the $\Gamma$-X and $\Gamma$-Y axes are protected by the M$_z$ symmetry. Due to the strong tilt along the $\Gamma$-X direction, the crossing point is type-II Dirac point. The crossing point on the $\Gamma$-Y axis is not tilted along the $\Gamma$-Y direction and is therefore type-I Dirac point. Since the band crossing points on the different directions can lead to a nodal ring, CrB$_4$ has a nodal ring with both type-I and type-II Dirac points protected by the M$_z$ symmetry. In addition, there is another band crossing in the $\Gamma$-X direction protected by M$_y$(1/2, 1/2, 1/2) symmetry, which is marked by a red arrow. We guess that there may be two small nodal rings in the k$_y$=0 plane protected by the M$_y$(1/2, 1/2, 1/2) symmetry. To prove our guess, we calculate all the nodes for the blue and yellow bands in the entire BZ, which is shown in Fig. 2(c). As guessed, CrB$_4$ has indeed two small nodal rings in the k$_y$=0 plane protected by M$_y$(1/2, 1/2, 1/2) symmetry (Fig. 2(c)). Since the two small nodal rings have a crossing with a larger nodal ring (Fig. 2(c)), CrB$_4$ is a topological nodal chain semimetal. Besides the nodal chain, there are also two nodal rings located in the non-high-symmetry surface of the BZ, which can be connected by the mirror M$_z$. The two nodal rings located in the non-high-symmetry surface of BZ are only protected by the IT symmetry (the T represents the time-reversal symmetry and I represents the space-reversal symmetry). In order to show the nodal ring of CrB$_4$ more directly, we also calculate the three-dimensional diagram of the yellow and blue bands. From Fig. 2(b), the nodal ring in the M$_z$=0 plane has relatively large dispersion and the type-I Dirac points below the Fermi levl are clearly shown, which is conducive to observation by the ARPES experiment. Therefore, CrB$_4$ is an ideal Dirac nodal chain topological semimetal.

Then, we also calculate the three-dimensional Fermi surface of the CrB$_4$, which is shown in Fig. 2(d). From Fig. 2(d), CrB$_4$ has a hole-type Fermi surface (blue) and an electron-type Fermi surface (yellow). Moreover, our calculations show that the volume of the hole Fermi surface is equal to that of the electron Fermi surface, which results in the compensation of the hole and electron carriers\cite{compensation,magnetoresistant-4}. Meanwhile, the nodal chain and nodal rings around the Fermi level make the hole and electron carriers have high mobility. According to the two-band model, CrB$_4$ has a large magnetoresistance effect when the applied electric and magnetic fields are perpendicular to each other. Due to the existence of type-II Dirac fermions, CrB$_4$ may have anisotropic negative magnetoresistance effect when the applied electric and magnetic fields are parallel\cite{Type-II, Duanprb}.

\begin{figure}
\centering
\includegraphics[width=1\linewidth]{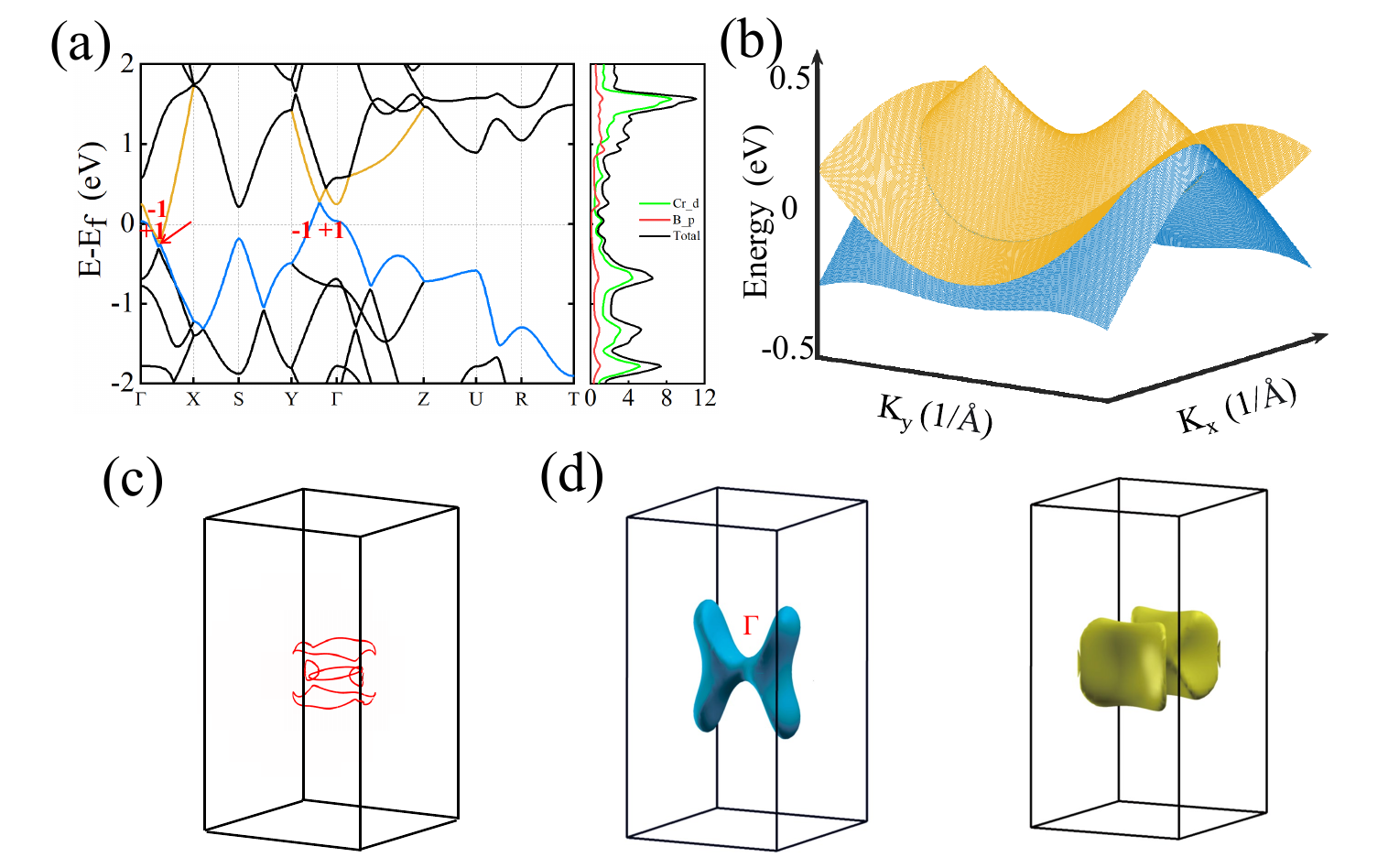}
\caption{\label{fig:2}(a) The band structure along high-symmetry directions(left panel) and density of states (right panel) for CrB$_4$ without SOC. The +1 and -1 are the eigenvalues of the mirror M$_z$. (b). Bulk energy dispersions of highest valence (bule) and lowest conduction (yellow) bands in the k$_z$= 0 plane. (c) The nodal rings and chain in the BZ for CrB$_4$. (d) The hole-type (left panel) and electron-type (right panel) Fermi surfaces.}
\end{figure}

It is well known that nodal ring semimetals have topologically protected boundary states. CrB$_4$ has five nodal rings which can be projected onto the (001) and (010) surfaces (Fig. 2(c)). Therefore, CrB$_4$ has topologically protected boundary states on the both (001) and (010) surfaces. Then, we calculate the boundary states of CrB$_4$ on the (001) and (010) surfaces, which are shown in Fig. 3 and Fig. 4, respectively. For the (001) surface, CrB$_4$ has three topologically protected surface states crossing  the Fermi levels deriving from the three nodal rings projected onto the (001) surface. More importantly, the Fermi arcs of CrB$_4$ at the Fermi level span the entire BZ. For the (010) surface, CrB$_4$ has two topologically protected surface states near the Fermi level deriving from two small nodal rings projected onto the (010) surface. Moreover, the Fermi arcs at the Fermi level also span the entire BZ. The Fermi arcs across the entire BZ at the Fermi level may result in CrB$_4$ having good catalytic property\cite{Li-apl, xie-sm}.


\begin{figure}
\centering 
\includegraphics[width=1\linewidth]{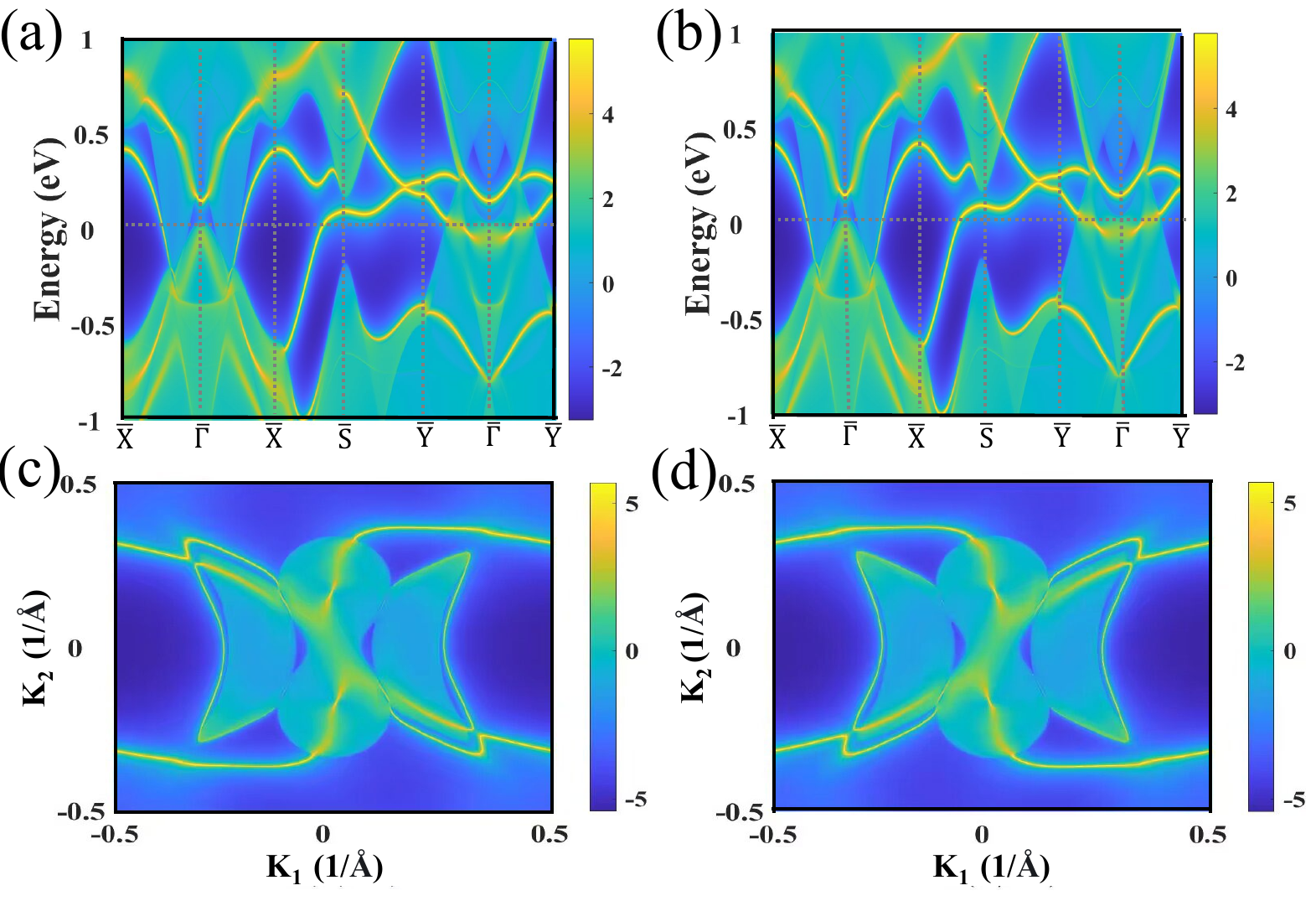}
\caption{\label{fig:3} Spectral function along the high-symmetry directions of CrB$_4$ left (a) and right (b) terminations for the (001) surface, which are contributed by Cr-B layers. The Fermi arcs at the Fermi level for left (c) and right (d) terminations.}
\end{figure}

\begin{figure}
\centering 
\includegraphics[width=1\linewidth]{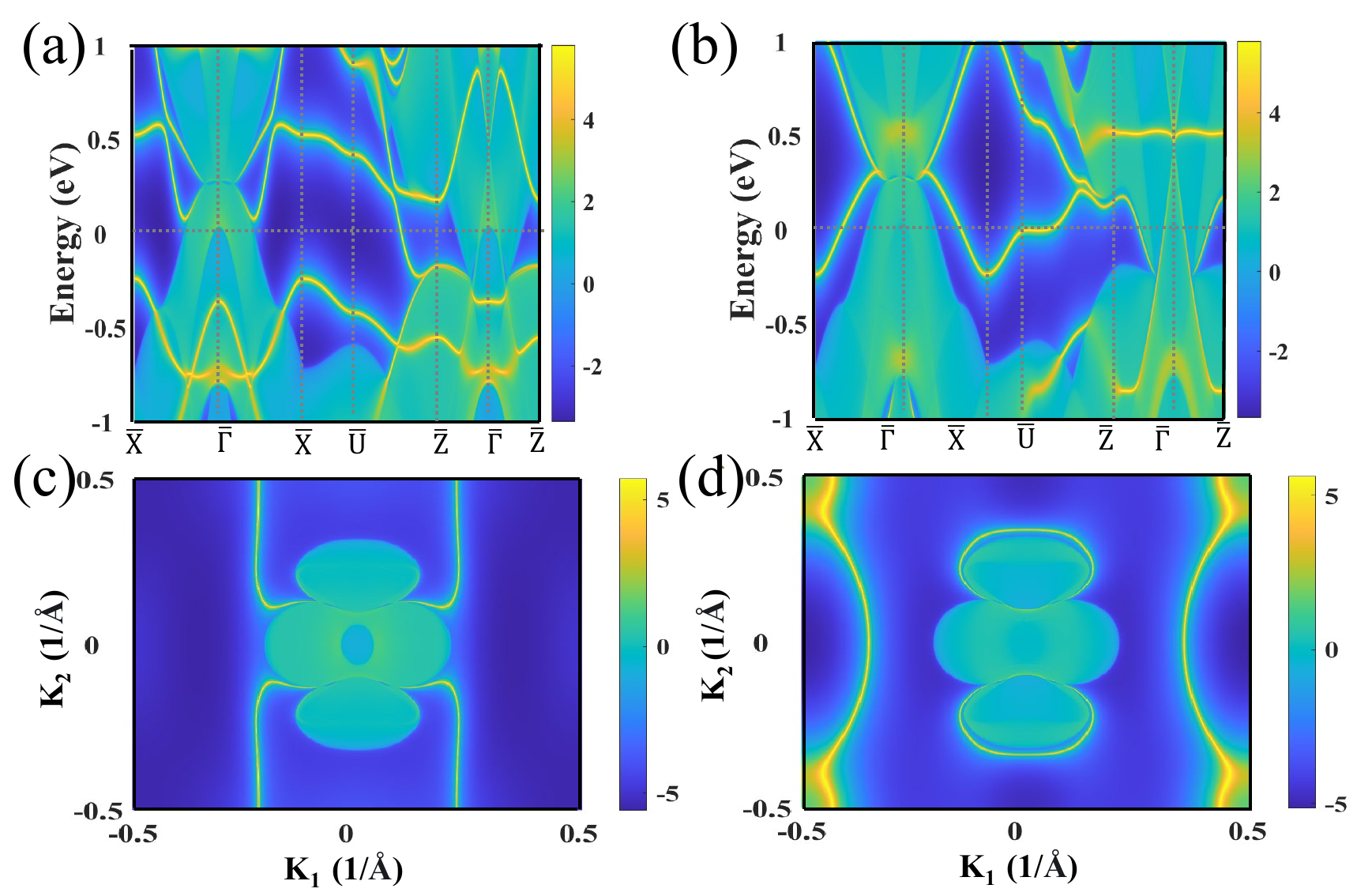}
\caption{\label{fig:4} Spectral function along the high-symmetry directions of CrB$_4$ left (a) and right (b) terminations for the (010) surface, which are contributed by B layers. The Fermi arcs at the Fermi level for left (c) and right (d) terminations.}
\end{figure}

When considering SOC, the symmetry of CrB$_4$ changes from D$_{2h}$ point group to D$_{2h}$ double point group. The highest symmetry for the high-symmetry axis in the BZ is C$_{2v}$ double group, such as the $\Gamma$-X axis. Since there is only one two-dimensional irreducible representation for C$_{2v}$ double group, both the nodal rings and nodal chain of CrB$_4$ must open an energy gap. Since both B and Cr atoms are light elements, the SOC effect of CrB$_4$ is very weak. With SOC(as shown in Fig.6), the sizes of the open bandgaps at the $\Gamma$-X and $\Gamma$-Y Dirac points are 2 meV, 27meV, and 18 meV respectively for CrB$_4$ (Fig. 5(a)). Considering that the SOC effect of the 2p orbitals of a B atom is much smaller than 1 meV, the four bands near the Fermi level are mainly contributed by the 3d orbitals of the Cr atom, which is also proved by the calculation of the DOS of CrB$_4$ (Fig. 2(a)). Due to the weak SOC effect, CrB$_4$ can be a model material for studying the novel physical properties of the type-II Dirac nodal chain in experiment.

\begin{table}
\centering
\caption{The product of parity of eight time-reversal-invariant points (TRTPs) for the 24 bands below the yellow band.}
\label{I}
\begin{tabular}{l l l l l l l l l l}
\hline
TRIPs   & $\Gamma$ & Z & U & R & T & X & Y & S & Total \\
\hline
parity & - & + & + & + & + & + & + & + &    - \\
\hline
\end{tabular}
\end{table}

On the other hand, what topological phase will CrB$_4$ transition from Dirac nodal chain semi-metal to under SOC $?$ Since the parities of the two inversion bands at the time-reversal invariant $\Gamma$ point are opposite (Fig. 5(a)), CrB$_4$ may be a strong topological insulator. In order to determine the Z$_2$ topological invariants of CrB$_4$, we need to calculate the parities of the eight time-reversal invariant points for 48 bands below the yellow bands\cite{Z2}.  

\begin{small}
\begin{align}
(x,y,z)&\xrightarrow{C_{2x}(\frac{1}{2}, \frac{1}{2}, \frac{1}{2})}(x+\frac{1}{2},-y+\frac{1}{2},-z+\frac{1}{2})\nonumber\\
&\xrightarrow{I}(-x-\frac{1}{2},y-\frac{1}{2},z-\frac{1}{2})
\end{align}
\end{small}

\begin{small}
\begin{align}
(x,y,z)&\xrightarrow{I}(-x,-y,-z)\nonumber\\
&\xrightarrow{C_{2x}(\frac{1}{2}, \frac{1}{2}, \frac{1}{2})}(-x+\frac{1}{2},y+\frac{1}{2},z+\frac{1}{2})
\end{align}
\end{small}


In fact, the nonsymmorphic symmetry of CrB$_4$ can greatly simplify the calculation of Z$_2$ topological invariants. The SOC effect does not change the parity of time-reversal invariant points, thus we only need to perform symmetry analysis without SOC. According to equations (1) and (2), C$_{2x}$(1/2, 1/2, 1/2)I=IC$_{2x}$(1/2, 1/2, 1/2)e$^{-ikT}$, where T represents the translation (1,1,1). Obviously, the fractional translation (1/2, 1/2, 1/2) makes C$_{2x}$ and I anticommutative at the time-reversal invariant X, Y, Z, and R points. Suppose $\Psi$ is an eigenstate of I with eigenvalue to be 1. Since C$_{2x}$(1/2, 1/2, 1/2) and I are anticommutative, C$_{2x}$(1/2, 1/2, 1/2)$\Psi$ is eigenstate of I with eigenvalue to be -1. Moreover, the Hamiltonian of CrB$_4$ commutes with C$_{2x}$(1/2, 1/2, 1/2) and I, so $\Psi$ and C$_{2x}$(1/2, 1/2, 1/2)$\Psi$ form a double degenerate state at time-reversal invariant X, Y, Z, and R points as shown in Fig. 2(a). Since the symmetry analysis is general, every band is of double degeneracy at the time-reversal invariant X, Y, Z, and R points, and the parity product of every doubly degenerate band is -1. There are 12 doubly degenerate bands below the yellow band, so the parity product of the occupying states at the time-reversal invariant points X, Y, Z and R is 1. 

Similar to the time-reversal invariant X Y, Z and R points, our analysis indicates the C$_{2x}$(1/2, 1/2, 1/2) and M$_z$ being anticommutative at the time-reversal invariant U and T points. If $\Psi$ is the eigenstate of M$_z$ with the eigenvalue 1, C$_{2x}$(1/2, 1/2, 1/2)$\Psi$ is the eigenstate of M$_z$ with the eigenvalue -1. The Hamiltonian of CrB$_4$ is commutative with both C$_{2x}$(1/2, 1/2, 1/2) and M$_z$, so $\Psi$ and C$_{2x}$(1/2, 1/2, 1/2)$\Psi$ form a doubly degenerate state. Given that I is also commutative with both C$_{2x}$(1/2, 1/2, 1/2) and M$_z$, $\Psi$ and C$_{2x}$(1/2, 1/2, 1/2)$\Psi$ have the same parity. Since every band is of double degeneracy at the time-reversal invariant U, and T points, and the parity product of every doubly degenerate band is 1, so the product of the parity of the time-reversal invariant U and T points is 1 for the 12 doubly degenerate bands below the yellow band. 

For the time-reversal invariant S point, $(C_{2x}(1/2, 1/2, 1/2)T)^2$ is equal to -1, which causes $\Psi$ and C$_{2x}$(1/2, 1/2, 1/2)$T\Psi$ to become Kramers degenerate. Since C$_{2x}$(1/2, 1/2, 1/2)T and I are commutative, the product of parity of every degenerate state at the time-reversal invariant S point is 1. Correspondingly, the product of parity at the time-reversal invariant S point is also 1 for the  12 doubly degenerate bands below the yellow band. These results of the above symmetry analysis are shown in Table 1. Thus the Z$_2$ topological invariant of CrB$_4$ depends only on the parity of the $\Gamma$ point. Our calculations show that the product of parity at the time-reversal invariant $\Gamma$ point for the 24 non-degenerate bands below the yellow band is -1 (table I). When SOC is considered, the 24 bands below the yellow band become 48 bands. However, the calculation of the Z$_2$ topological invariants requires only the product of the parity of half of the 48 bands, which is consistent with the result of the parity product of the 24 bands without SOC. According to these results in Table I, the Z$_2$ topological invariants are (1,000) for CrB$_4$, which is also consistent with calculations of WannierTools packages. Therefore, CrB$_4$ is a strong topological insulator. 

The strong topological insulator has nontrivial Dirac surface states protected by the time-reversal symmetry. The calculated topological surface states of CrB$_4$ on the (001) surface are shown Fig. 5(b). Compared with Fig. 3(a), one topologically protected surface state splits into two, but the splits are very small, so the Dirac points of the surface state protected by the time-reversal symmetry are not obvious, which also implies the weak SOC effect of CrB$_4$. Since there are only two non-trivial bands around the Fermi level and its topologically protected surface states at the Fermi level span the entire BZ, CrB$_4$ is an ideal type-II Dirac nodal chain semimetal with distinct properties.

\begin{figure}
\centering
\includegraphics[width=1\linewidth]{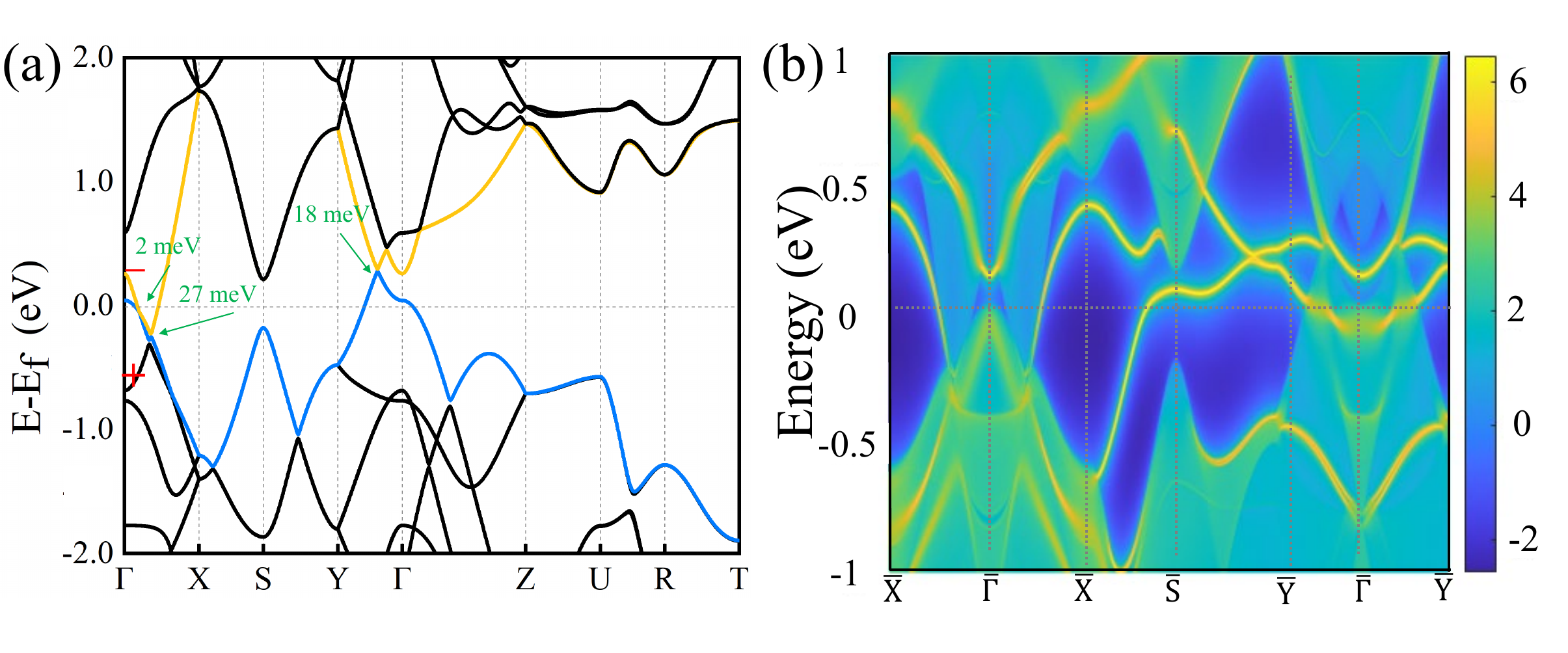}
\caption{\label{fig:5} (a) The band structure for CrB$_4$ with SOC along high-symmetry directions. (b) Spectral function along the high-symmetry directions left terminations for the (001) surface. The '+' and '-' represent parity for space-inversion symmetry.}
\end{figure}

\section{In summary}

 Based on symmetry analysis and the first-principles electronic structure calculations, we predict that CrB$_4$ is a topological semimetal with a type-II nodal chain and two nodal rings protected respectively by mirror symmetry and IT symmetry. More importantly, the topologically protected drumhead surface states of CrB$_4$ span the entire BZ at the Fermi level. Meanwhile, there are only two nontrivial bands crossing the Fermi level, thus CrB$_4$ is an ideal type-II Dirac nodal chain semimetal with distinct properties. When considering SOC, CrB$_4$ transforms from topological semimetal phase to topological insulator phase. Because it has been synthesized experimentally and has a very weak SOC effect, CrB$_4$ is an ideal material platform for studying the exotic properties of type-II Dirac nodal chain semimetals in experiment.

\begin{acknowledgments}
 This work was financially supported by the National Natural Science Foundation of China (Grant No.11934020, No.12204533, No.62476278, No.62206299), the Fundamental Research Funds for the Central Universities, and the Research Funds of Renmin University of China (Grant No. 24XNKJ15) and the Innovation Program for Quantum Science and Technology (Grant No. 2021ZD0302402). Computational resources have been provided by the Physical Laboratory of High Performance Computing at Renmin University of China.
\end{acknowledgments}

\nocite{*}

\bibliography{ref}
\end{document}